# Enhancement of electromagnetically induced transparency in metamaterials using long range coupling mediated by a hyperbolic material

**ZHIWEI GUO,**[1,3] **HAITAO JIANG,**[1,3] **YUNHUI LI,**[1] **HONG CHEN,**[1,*] **AND G. S. AGARWAL**[2]

[1] *MOE Key Laboratory of Advanced Micro-Structured Materials, School of Physics Science and Engineering, Tongji University, Shanghai, 200092, China*
[2] *Institute for Quantum Science and Engineering and Department of Biological and Agricultural Engineering, Texas A&M University, College Station, TX 77843, USA*
[3]*These authors contributed equally to this work*
*\* Corresponding author:hongchen@tongji.edu.cn*

**Abstract:** Near-field coupling is a fundamental physical effect, which plays an important role in the establishment of classical analog of electromagnetically induced transparency (EIT). However, in a normal environment the coupling length between the bright and dark artificial atoms is very short and far less than one wavelength, owing to the exponentially decaying property of near fields. In this work, we report the realization of a long range EIT, by using a hyperbolic metamaterial (HMM) which can convert the near fields into high-$k$ propagating waves to overcome the problem of weak coupling at long distance. Both simulation and experiment show that the coupling length can be enhanced by nearly two orders of magnitude with the aid of a HMM. This long range EIT might be very useful in a variety of applications including sensors, detectors, switch, long-range energy transfer, etc.

## 1. Introduction

The hyperbolic metamaterials (HMMs) have attracted considerable interest. These materials can mimic some of the properties of negative-index materials such as negative refraction [1-4]. It is also possible to tailor such materials by going from elliptic to hyperbolic dispersion relations [5-7]. The density of states in the vicinity of HMM can be large [8-10] and this has important consequences for spontaneous emission [11, 12]. Since HMM can allow very large surface vectors to propagate, unlike metals, these are finding applications in high-resolution imaging [13-18], unusual surface wave [19], cavity [20], scattering [21, 22], thermal radiation [23-25], and giant photoresponsivity [26]. Recently Biehs et al, Cortes and Jacob have discovered a very long range dipole-dipole interaction in a HMM [27-30]. This is especially significant in the energy transfer from a donor atom to an acceptor atom and can lead to significant entanglement between two atoms, which is important for quantum information tasks. All such studies demonstrate the importance of a HMM in a range of applications. Other interesting possibilities include coupling of macroscopic objects like nanoparticles or fundamental interaction between a dipole and a quadrupole.

We present a theoretical and experimental study of the coupling between a dipole-like bright atom and a quadrupole-like dark atom which is especially relevant in the classical analog of electromagnetically induced transparency (EIT) based on metamaterials. The realization of the EIT which derives the interference-based high-$Q$ resonance inside the stop band has attracted much attention in terms of various applications: slow light, sensors, switch, etc [31-37]. We note that the EIT in meta systems like three-level system is well known. This requires significant coupling between the bright atom and dark atom modes. Thus the distance between the bright and dark atoms has to be $\lambda/20 \sim \lambda/40$, where $\lambda$ is the operation wavelength. For larger distances this coupling is insignificant and the dark atom remains dark when the bright atom is excited. A fundamental question is then how the coupling can be enhanced. One very good possibility is to use the recently discovered long range coupling effects mediated by the HMM. Inspired by this mechanism, we show that with the aid of a HMM, EIT can also be maintained even when the separation between bright and dark atoms is large.

In this work, we will demonstrate the enhancement effect of HMM for the long range EIT by comparing short distance one in a normal material. In a normal material background such as air, EIT realized by a three-level system does not occur when the separation of bright and dark atoms is very large. This is because the iso-frequency contour (IFC) of the normal material is a closed surface and the high-$k$ modes are excluded. However, for a HMM, it supports large $k$ modes because of the open hyperboloid dispersion [38]. The evanescent fields existing in the normal material can be converted to the high-$k$ propagating waves in the HMM. By using epsilon-near-pole (ENP) HMM which has been validated to have a large dipole-dipole interaction [27], we attempt to achieve a long range EIT. Although the long range coupling occurs for any kind of HMM, the ENP-HMM is the most efficient one (see Appendix $A$ for details). Our experimental work demonstrates that an introduced HMM can greatly enhance the interaction distance between a dipole-like bright atom and a quadrupole-like dark atom previously governed by the near-field coupling.

The paper is organized as follows. At first, we consider a general theoretical model of long range EIT that considers the near-field coupling and the converted-far-field coupling between bright and dark atoms, simultaneously. Then, based on microwave transmission-line systems, we perform both simulations and experiments to check the validity of the model and analyze the physical mechanism of the long range EIT. If the model only has near-field coupling term as in the case of a normal medium, the EIT will disappear when the distance between the bright and dark atoms is large. We then put a HMM in the intervening space of the two atoms and show how EIT is restored. In this long range situation, the establishment of EIT mainly comes from the high-$k$ propagating mode because the near-field coupling is very weak. Combining simulation and experimental results, we show that the HMM can enhance the coupling distance by nearly two orders of magnitude in contrast to the situation in a normal environment. This HMM-mediated long range coupling may provide inspiration for the design of new EIT platforms relevant to sensing and detection.

## 2. Theoretical model and analysis

Firstly, we establish a physical model of long range EIT in a three-level system, as is schematically shown in Fig. 1. In this model, both near-field coupling between bright and dark atoms contributed by the ordinary mode in a HMM and the converted-far-field coupling contributed by the extraordinary modes are considered. In Fig. 1, the lower red sphere can be directly excited by the incoming wave $\tilde{S}_{in}$, which could serve as a bright atom. The upper green sphere cannot be directly excited by the incoming wave and it need be excited via the bright atom. Based on the coupled mode theory, the time-harmonic energy amplitudes for two resonance modes can be defined as $\tilde{a}_1 = a_1 e^{i\omega t}$ and $\tilde{a}_2 = a_2 e^{i\omega t}$, where $\omega$ is the angular frequency. When an incident wave $\tilde{S}_{in} = S_{in} \cdot e^{i\omega t}$ launches into the structure, the dynamic equations are given by

$$\frac{d\tilde{a}_1}{dt} = (i\omega_1 - \Gamma_1 - \gamma_1)\tilde{a}_1 + i\sqrt{\gamma_1}\tilde{S}_{in} + i(\kappa + i\sqrt{\gamma_1\gamma_2}e^{-i\varphi})\tilde{a}_2,$$
$$\frac{d\tilde{a}_2}{dt} = (i\omega_2 - \Gamma_2 - \gamma_2)\tilde{a}_2 + i(\kappa + i\sqrt{\gamma_1\gamma_2}e^{-i\varphi})\tilde{a}_1 \qquad (1)$$

where $\omega_1$ and $\omega_2$ are the center frequencies of two resonances, respectively. In Eq. (1), the terms $i\kappa\tilde{a}_1$ and $i\kappa\tilde{a}_2$ describe the near-field interaction between the two resonators while the term $i\sqrt{\gamma_1}\tilde{S}_{in}$ describes the coupling between resonator one and the input wave. The coupling term with phase $\varphi$ comes from the converted high-$k$ propagating modes in the HMM. The decay rate of each resonator includes radiative loss $\gamma_i (i=1,2)$ and dissipative loss $\Gamma_i (i=1,2)$. The transmittance (abbreviated as '$T$') of the system from the input port and to the output port can be derived as [39]

$$T = \left|1 + \frac{-\gamma_1[i(\omega-\omega_2)+\Gamma_2+\gamma_2]}{[i(\omega-\omega_1)+\Gamma_1+\gamma_1]\cdot[i(\omega-\omega_2)+\Gamma_2+\gamma_2]+(\kappa+i\sqrt{\gamma_1\gamma_2}e^{-i\varphi})^2}\right|^2. \qquad (2)$$

From Eq. (2), we would expect a Fabry-Perot type of oscillation in transmission with respect to the phase $\varphi$ in a period of 180 degrees if the near-field coupling term can be neglected ($\kappa \to 0$).

Then, based on microwave transmission lines (TLs), we perform both simulations and experiments to check the validity of the model. The effectiveness of using TLs to mimic artificial atoms has been verified in [40, 41]. In fact, in Eq. (1), if there is no coupling term with phase $\varphi$, that is, only

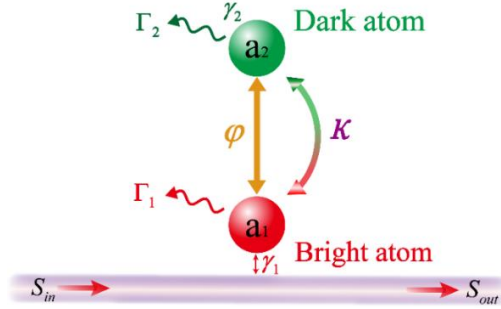

Fig. 1. Schematic of a long range EIT in a three-level system. This general model considers both near-field coupling and converted-far-field coupling between bright and dark atoms.

$$T = \left| 1 + \frac{-\gamma_1[i(\omega-\omega_2)+\Gamma_2+\gamma_2]}{[i(\omega-\omega_1)+\Gamma_1+\gamma_1]\cdot[i(\omega-\omega_2)+\Gamma_2+\gamma_2]+\kappa^2} \right|^2. \qquad (3)$$

the near-field coupling term connected with $\kappa$ exists, the model just describes the case of a conventional EIT in a normal medium. We firstly compare the results from the model and those from the simulations in this conventional EIT case. The three-level artificial atom system and the related parameters are given in Fig. 2(a). In this case, the transmittance $T$ of the system can be written asIn Fig. 2(a), the artificial atoms couple to the incident wave by the TL [40, 41]. The narrow split ring resonator (SRR) connected to the TL can be directly excited by the incident wave propagating in the TL and serves as a bright atom. In contrast, the wide SRR on the side of the narrow SRR acts as the dark atom [42], because it is far from the TL and need be excited through the narrow SRR by the near-field coupling between them. Moreover, to avoid detuning, we choose $C_1 = C_2 = 1$ pF to make the resonance frequencies of the two SRRs be the same. The geometric parameters $a_1, a_2, d, b, g$ are 4.4, 10.2, 4.2, 15, 0.8 mm, respectively. The metallic line width of the SRRs is 0.2 mm and the width of the TL is 2.8 mm. All of them are fabricated on a FR-4 ($\varepsilon_r = 4.35, \tan\theta = 0.03$) substrate with a thickness of $h = 1.6$ mm. We first simulate the transmittance spectra by the CST software. When the separation between two atoms is small, e.g., $s = 0.2$ mm, an EIT window around 0.79 GHz appears, as is shown by the purple scattered open dots in Fig. 2(b). However, when $s$ gradually increases, the EIT peak becomes narrower and narrower and at the same time shallower and shallower, as is shown in Fig. 2(b). When $s$ reaches a large value, the EIT disappears and the transmittance spectrum degenerates into that of the single bright atom. In addition to the simulated transmittance (shown by the scattered open dots), the calculated transmittance spectra based on Eq. (3) are also given in Fig. 2(b), as shown by the solid lines. Overall, the results from the model and the simulations are in good agreement. Moreover, based on Eq. (3), the outgoing transmittance spectra versus frequency and $s$ are calculated and shown by the background color in Fig. 2(c). The split frequencies from the CST simulation and the experiments are marked by the blue solid dots and red open dots, respectively. The coupling-induced frequency split will disappear as $s$ increases. In the calculations of Eq. (3), the value of $\kappa$ is obtained from the simulated transmittance spectrum. From the spectral response, we can fit a decaying exponential function of $\kappa$ as

$\kappa = 0.105 e^{-\frac{s}{0.99}}$, see Fig. 2(a). From the spectrum, we can also obtain $\gamma_1 = 0.017$ GHz, $\gamma_2 = 0.009$ GHz, $\Gamma_1 = 0.0045$ GHz, and $\Gamma_2 = 0.0047$ GHz. The value of $s$ and thereby $\kappa$ strongly influence the transmittance. We calculate the 2D electric field distributions at 0.79 GHz for two different $s$. For $s = 0.2$ mm, $\kappa$ is large and the electric fields are transferred from the bright atom to the dark atom, as is shown in Fig. 2(d). However, for $s = 5$ mm, $\kappa$ is very small and the electric fields are mainly concentrated at the bright atom, see Fig. 2(e). So, in a normal environment, the EIT no longer occurs for the large separation.

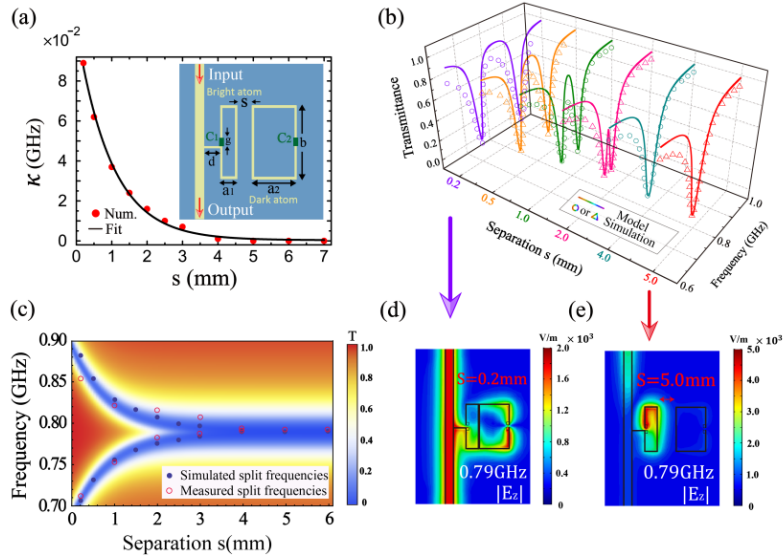

Fig. 2. Microwave transmission-line system. (a) The coupling strength decreases exponentially with the increase of distance. The structure and related parameters are shown in the inset. (b) Analytical calculated (solid lines) and simulated (scattered open dots) transmittance spectra as the function of $s$. (c) The background color gives analytical outgoing transmittance spectra $T$ for different $s$. The red open (blue solid) circles are the measured (simulated) split frequencies. (d, e) The 2D electric field distributions when $s = 0.2$ and 5 mm, respectively.

However, the coupling distance can be strongly enlarged if we add a HMM between the two atoms. Although both ordinary and extraordinary modes would be excited in the HMM [43], the near-field coupling contribution from the ordinary modes is negligible when the coupling distance is long, which is similar to the case of a normal material. For a short coupling distance, the near-field coupling contribution from the ordinary modes will play roles in the establishment of EIT. However, similar to [7], in our designed TLs loaded with lumped elements, the excited modes are mainly the transverse-electric (TE) polarized extraordinary modes. The impact of TM-polarized ordinary modes in our TL system can be neglected (see Appendix B for details). The schematic of the HMM and the related anisotropic 2D-circuit model are shown in Fig. 3(a). The width of TL (denoted by $w$) is 2.8 mm and the length of a unit cell (denoted by $p$) is 15 mm. In the circuit, the structure factor $g = [1.393 + w/h + 0.667\ln(w/h) + 1.444]^{-1} = 0.255$ when $w > h$. We load lumped capacitors with $C = 0.1$ pF in the $x$ direction to ensure a negative $\mu_y$ [44, 45]. The dispersion relation of TE-polarized waves in TLs is described by

$$\frac{k_x^2}{\mu_y} + \frac{k_y^2}{\mu_x} = \varepsilon \cdot k_0^2. \qquad (4)$$

In the long-wavelength limits, if we do not consider the loss, the effective permittivity and permeability of 2D TLs can be written as [7, 46]

$$\varepsilon = \frac{2C_0 \cdot g}{\varepsilon_0}, \mu_x = \frac{L_0}{g \cdot \mu_0},$$

$$\mu_y = \frac{L_0}{g \cdot \mu_0} - \frac{1}{\omega^2 \cdot C \cdot p \cdot g \cdot \mu_0}, \quad (5)$$

where $\varepsilon_0$ and $\mu_0$ are the permittivity and permeability of vacuum, respectively. $C_0$ and $L_0$ are the per-unit length inductance and capacitance of the TL, respectively. In conventional classical EIT, the dielectric or air background is lossless, while in HMM-mediated long range EIT, the HMM is lossy. So it is very necessary to study the influence of the loss of the HMM on the long range EIT. For the TL-based effective HMM, the loss mainly comes from the dielectric loss of substrate and the Ohm loss of the loaded capacitor in the $x$ direction. Considering the dielectric loss of substrate, the imaginary part of permittivity can be written as

$$\text{Im}\,\varepsilon = \frac{\text{Re}\,\varepsilon \cdot \text{Tan}\,\theta}{2} + \frac{\text{Re}\,\varepsilon \cdot \text{Tan}\,\theta}{2}(1 + 12\frac{h}{w})^{-\frac{1}{2}}. \quad (6)$$

Based on Eqs. (5) and (6) we can get $\varepsilon = 6.5 + 0.18i$, $\mu_x = 1$ and $\mu_y = 1 - \frac{2.08 \times 10^{21}}{\omega^2}$. The dependence of $\mu_y$ on the frequency is drawn in Fig. 3(b). At $\omega/2\pi = 0.8\,\text{GHz}$, $\mu_y \approx -81$. This value is negative enough to ensure the ENP-HMM, as are shown by the red solid line in Fig. 3(c). For comparison, a general hyperbolic dispersion when $\mu_y = -1$ is also given in Fig. 3(c), as are shown by the blue solid line in Fig. 3(c). In addition to the dielectric loss, considering the Ohm loss of loaded capacitors in the $x$ direction, we can calculate the imaginary part of permeability in the $y$ direction as

$$\text{Im}\,\mu_y = \frac{1}{\mu_0} \cdot \frac{R_c}{\omega p g}, \quad (7)$$

where $R_c$ denotes the Ohm loss of capacitors. However, after calculating the dispersion diagram of HMM, we find that $\text{Im}\,\mu_y$ will not introduce the imaginary part of $k_y$ (denoted by $\text{Im}\,k_y$). This means that $\text{Im}\,\mu_y$ will not introduce absorption in the propagating direction and will not influence the long range EIT. Therefore, in Figs. 3(b) and 3(c), we do not consider $\mathrm{Im}\,\mu_y$. From Fig. 3(c), we see that $\mathrm{Im}\,\varepsilon$ will introduce small values of $\mathrm{Im}\, k_y$, as are shown by the dashed lines. The appearance of $\mathrm{Im}\, k_y$ will increase the absorption and decrease the transmission window of the EIT. Besides considering the effect of loss, we need consider the influence of the type of HMM on the long range EIT. Although both type I and type II HMM can convert the near fields into propagating waves, the direction of the converted propagating wave is determined by the designed HMM. For example, in our system, to make the converted-far-field propagate along the $y$ direction, a type I HMM [16] is needed. If the type of HMM is changed, the direction of the converted propagating wave would change and the wave would not interact well with the dark atom. In this case the EIT would be hard to establish.

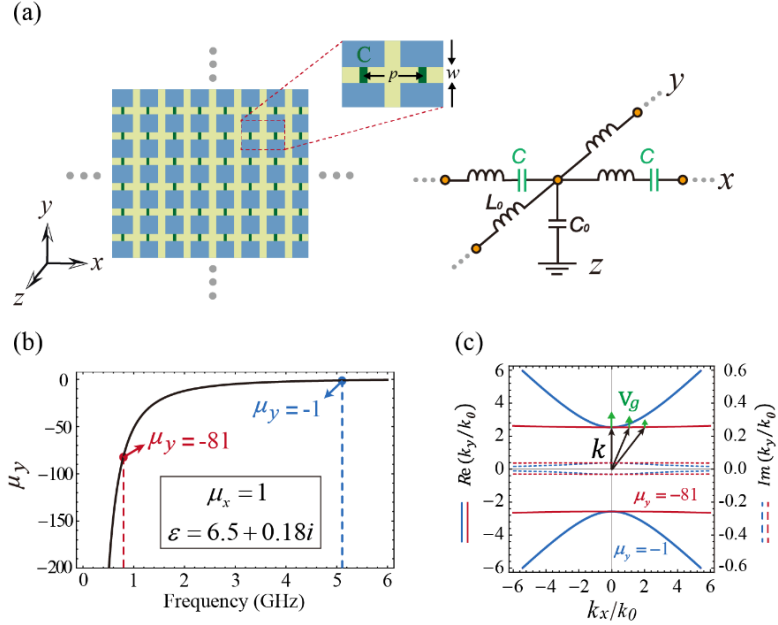

Fig. 3. (a) The structure and the related anisotropic 2D-circuit model of microwave HMM. (b) The effective parameters based on the TLs. $\mu_y = -81$ and $\mu_y = -1$ are marked by the red and blue dots, respectively. (c) The IFCs of the ENP-HMM are two flat lines (red line). For comparison, the IFCs of a general HMM are also given (blue line). The wave vectors (group velocity) are shown by the black (green) arrows. For the complex dispersions, $\mathrm{Re}\, k_y$ ($\mathrm{Im}\, k_y$) is shown by solid (dashed) line.

### 3. Simulations and experiments

Now we add the TL-based effective HMM between the bright and dark artificial atoms. The schematic of the structure to realize the long range EIT is shown in Fig. 4. The effective HMM is added between the two SRRs whose geometric parameters are the same as those in Fig. 2(a). The effective HMM contains 4 (6) unit cells in the $x$ ($y$) direction. The parameters of the two atoms remain unchanged except that $C_2 = 1.2$ pF. Moreover, for convenience, the dark atom is put near a single TL out of the effective HMM. In this sample, the distance between the two atoms is 60 mm which is much longer than the effective coupling length in a normal environment in Fig. 2(a). In the HMM, the phase of the converted propagating wave is equal to the product of the wavevector in the propagating direction and the distance ($s$). For an ENP-HMM, the dispersions are flat lines and the fields are collimated in the y direction. In this case, the converted-far-field phase $\varphi$ in HMM can be written as

$$\varphi = k_y s = \sqrt{\varepsilon \mu_x} k_0 s, \tag{8}$$

where $k_0$ is the wave vector of free space. From Eq. (8), we find again that only the imaginary part of $\varepsilon$ will lead to the imaginary part of $\varphi$ and thereby introduce the absorption. Moreover, in our system, in addition to the converted-far-field phase $\varphi$ induced by the HMM, there is an additional propagating phase $\Delta\varphi$ from the finite size of the atoms and the added transmission line for dark atom coupling. In the CST simulation, we put four probes at the center of the bright atom, the left edge and the right edge of the HMM as well as the center of the dark atom, simultaneously. Then, we extract the phase difference from the center of the bright atom to the left edge of the HMM and that from the right edge of the HMM to the center of the dark atom. By adding the two phase differences, we obtain $\Delta\varphi = 137^o$. Of course, the probe method can only extract the real part of $\Delta\varphi$. Nevertheless, the imaginary part of $\Delta\varphi$ induced by the dielectric loss of the substrate at the position of the bright atom and the dark atom has been considered in the dissipative loss $\Gamma_i (i=1,2)$ of the two atoms. Therefore, to compare with the simulations, in the calculation of Eq. (2), the value of phase is modified to $\varphi + 137^o$, where $\varphi$ is a complex value obtained from Eq. (8). Now we compare the results from the model and those from the simulations in the case of HMM-mediated EIT. The calculated reflectance $R$ based on Eq. (1) can be derived as

$$R = (1 - \sqrt{T})^2, \tag{9}$$

where $T$ is given by Eq. (2). In the model, if we need not consider the contribution of ordinary modes, we will not consider the near-field coupling term connected with $\kappa$. In the simulations, $T$ and $R$ can be directly extracted. In all cases, the absorbance $A = 1 - R - T$. The spectra of $R$, $T$ and $A$ from the model and from the simulations for $s = 45$, 60 and 75 mm are given in the left column (dashed lines) and the right column (solid lines) of Fig. 5, respectively. When the distance $s = 60$ mm ($\kappa \to 0$), an EIT window is still clearly seen. In comparison with Fig. 2(a), the effective coupling distance is boosted by nearly two orders of magnitude. In the case of $s = 60$ mm, $\varphi_{s=60mm} = 148^o$.

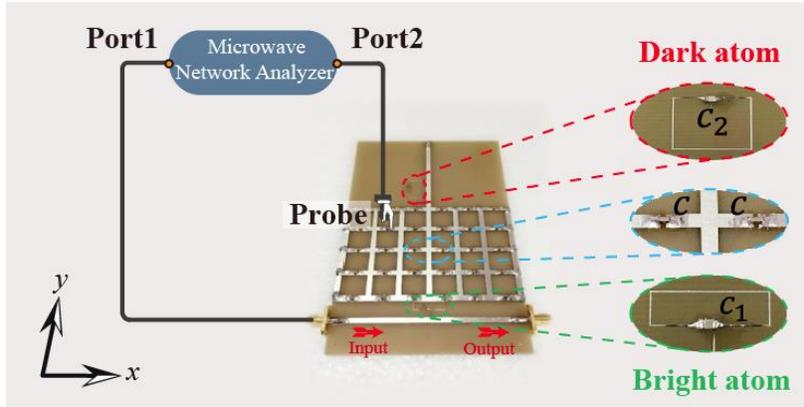

Fig. 4. Schematic of the structure to realize a long range EIT. Bright atom, dark atom and the unit of HMM are enlarged in the left, respectively.

For different $s$, $\varphi$ will be different and the corresponding spectra will change according to Eq. (2). It is seen from Fig. 5 that the spectrum changes in the cases of $s = 45$ mm and 75 mm, in which $\varphi_{s=45mm} = 111^o$ and $\varphi_{s=75mm} = 185^o$, respectively. Overall, the results from the model agree well with those from the simulations. Therefore, we demonstrate that the long range EIT can be well described by the model of Eq. (1).

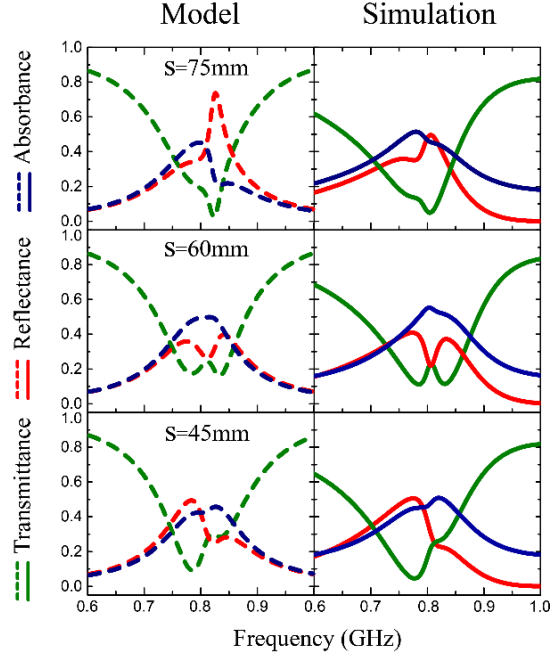

Fig. 5.The spectra of $R$, $T$ and $A$ from the model (dashed lines) and from the simulations (solid lines).With the aid of HMM, EIT can be well reestablished when the distance of bright and dark atoms is much larger than the normal near-field coupling distance. The EIT spectra is still seen when the distance between two atoms reaches 60 mm.

To emphasize the role of the HMM, we show a systematic comparison of a system that does show the effect (long-range EIT) with the system that does not show it. Transmission line systems with different values of capacitors are used for the working and non-working systems. In Fig. 6(a), the transmission line with $C = 0.1$ pF mimics the ENP-HMM. The simulated transmission spectra of this system when the distance $s = 60$ and 135 mm, respectively, are shown in Fig. 6(b). The EIT windows are clearly seen for two distances. In Fig. 6(c) and 6(d), we numerically show the electric field distributions corresponding to the EIT peaks in the cases of $s = 60$ and 135 mm, respectively. In two cases, the fields are transferred from the bright atom to the dark atoms, which demonstrate the long-range EIT. For comparison, we only change the value of $C$ from 0.1 pF to 30 pF. As a result, $\mu_y = 0.73$ and the transmission line mimics a normal material with a nearly circular dispersion, as is shown in Fig. 6(e). The transmission spectra of this system when $s = 60$ and 135 mm, respectively, are given in Fig. 6(f). There are no EIT windows in the opaque region. Moreover, the electric fields corresponding to the transmission dips in two cases of different distance are mainly confined in the bight atom, as are seen in Fig. 6(g) and 6(h). Therefore, the comparison between working and non-working systems in Fig. 6 convincingly shows that, with the aid of HMM, long-range EIT can be realized.

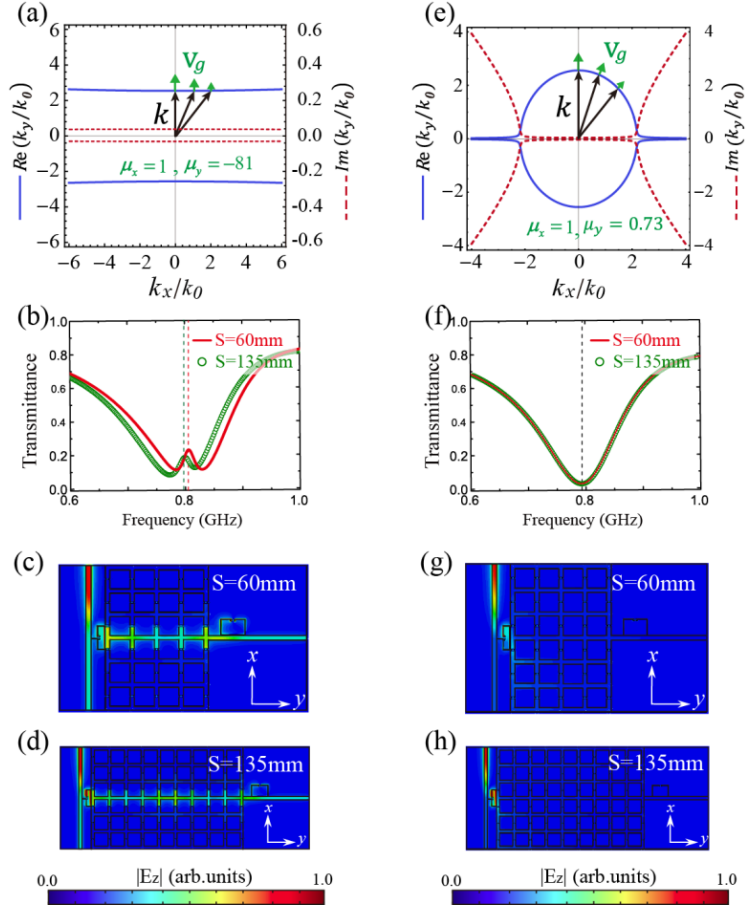

Fig. 6. The system (left column) with long-range EIT and that (right column) without long-range EIT. (a) The transmission line with $C=0.1$ pF mimics the ENP-HMM. (b), (f) The transmission spectra when $s=60$ and 135 mm, respectively. (e) The transmission line with $C=30$ pF mimics a normal material. (c), (d) The field distributions corresponding to the EIT peaks in (b). (g), (h) The field distributions corresponding to the transmission dips in (f).

To see the formation of the long range EIT clearly, now we experimentally study the change of field pattern from the case of a single bright atom to the case of a molecular with bright and dark atoms. For the bright atom, a Lorenz line shape can be seen from the reflectance spectra in Fig. 7(a). The corresponding energy will be collimated in the $y$ direction in the HMM, as are shown by the field distribution in Fig. 7(b) at the resonance frequency of 0.806 GHz. However, for the molecule with bright and dark atoms, a transparent EIT window (marked by the black dashed line) occurs in the reflectance spectra in Fig. 7(c). At this EIT frequency, the energy is transferred from the bright atom to the dark atom, see the measured electric field patterns in Fig. 7(d). In our experiments, the samples are placed on an automatic translation device which makes it feasible and accurate to probe the field distribution using a near-field scanning measurement. To measure the electric fields, the signal launches from the port one of vector network analyzer (Agilent PNA Network Analyzer N5222A) and another antenna (near-field probe) connecting to the port 2 of analyzer are employed to records the electric field pattern. The length of the rod antenna is 1 mm. It is vertically placed 1mm above the TLs to measure the signals of electric fields of the TLs in the 2D plane. The spatial step of scanning the near field is set to be 1 mm in the $x$ and $y$ directions, respectively. The field amplitudes are normalized according to their respective maximum amplitude. Comparing Fig. 5 in the case of $s = 60$ mm with Fig. 7(c), we see that on the whole the experimental results agree well with the simulated one. The deviations between the simulations and the experiments originate from the discrepancies of material parameters including the dielectric constant of the FR4 substrate and the value of capacitors between the theoretical data and the real fabricated sample.

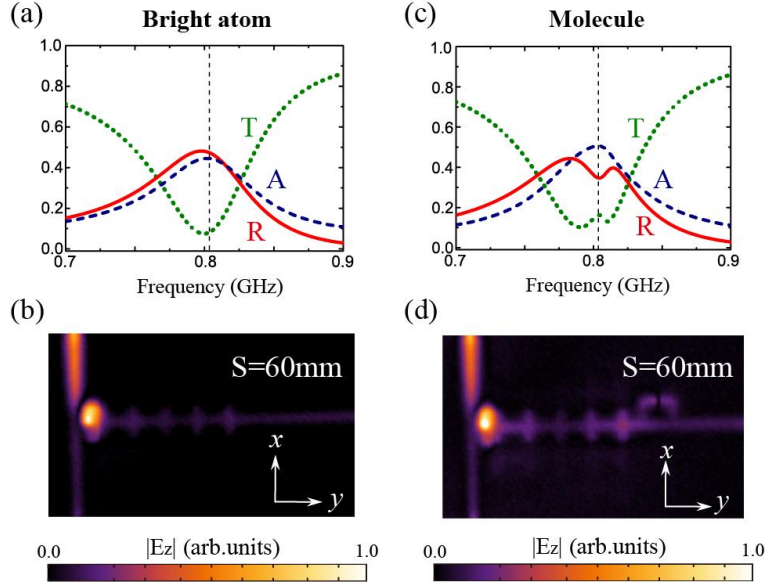

Fig. 7. The measured transmittance, reflectance and absorbance spectra of bright atom (a) and a molecule (c), respectively. The measured 2D electric field distributions of bright atom (b) and a molecule (d) at 0.806 GHz, respectively [marked by a dashed line in (a) or (c)].

Furthermore, to see the dependence of EIT on the separation $s$ and compare with Fig. 2(c) in the case of a conventional EIT, we plot Fig. 8 in the case of a HMM-mediated EIT. As is seen from Eq. (2), the transmittance depends on the converted-far-field phase $\varphi$ in the HMM. So at first we give the dependence of $\varphi$ on $s$. The linear relationship between the real part of $\varphi$ and $s$ is shown by the red solid line in Fig. 8(a), as is dictated by Eq. (8). For comparison, the simulated real part of $\varphi$ by extracting the phase difference from the left edge to the right edge of the HMM is also given in Fig. 8(a), as are shown by the scattered stars. It is seen that the scattered stars are nearly located at the red solid line. The dependence of the imaginary part of $\varphi$ on $s$ is also given in Fig. 8(a), as are shown by the black open circles. The imaginary part of $\varphi$ only increases a little bit when $s$ increases from 60 to 135 mm. This is because the imaginary part of $k_y$ is much smaller than the real part of $k_y$, see Fig. 3(c). As a result, in Fig. 6(b) the EIT window only decreases a little bit when $s$ changes from 60 to 135 mm. In Fig. 8(b) the background color is derived from Eq. (2) when $\kappa \to 0$. One can see that the cases of $s = 45$, 60 and 75 mm are very similar to $s = 120$, 135 and 150 mm, respectively, in which the phase difference is near 180 degrees. This is the Fabry-Perot type of oscillation in transmittance, as is indicated by Eq. (2). The scattered circles correspond to three structures in which $s = 45$, 60 and 75 mm, respectively. The measured (simulated) EIT split frequencies are marked by red open (black solid) circles. Overall, all the simulations and experiments are in good agreement and they demonstrate that the HMM-mediated EIT survives even when the coupling distance is far beyond the effective coupling length in a normal environment.

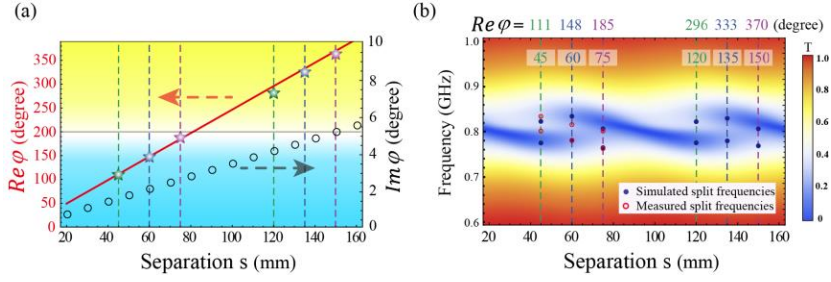

Fig. 8. (a) The dependence of $\mathrm{Re}(\varphi)$ (red solid line) and $\mathrm{Im}(\varphi)$ (black open circles) on $s$. (b) The dependence of HMM-mediated EIT on $s$. The background color gives analytical outgoing transmittance spectra $T$ for different $s$. Different dashed color lines denote different separations. The red open (blue solid) circles are the measured (simulated) EIT split frequencies.

## 4. Conclusion

As a summary, the phenomenon of long range EIT observed in our experiments confirms the theoretical prediction that the energy transfer will be enhanced strongly in a HMM environment. This experimental work demonstrates that the HMM can overcome the short distance limitation of near-field coupling in a conventional medium. The HMM-enhanced coupling would have profound influences on the physical processes previously governed by the near-field coupling. Moreover, this long range EIT might be very useful in various applications including slow light, sensors, switch, long-range energy transfer, etc.

## Appendix

### A. Long range EIT can be realized in a general HMM

Although an ENP-HMM is used in the main text, we emphasize that the long range coupling can occur for any kind of HMM. Here we study the EIT mediated by a general HMM. To obtain a general HMM, we change the lumped capacitance in Fig. 4 into 2.5 pF. Then we derive the effective parameters of TLs based on Eq. (5), as are shown in Fig. 9(a). At the reference frequency of 0.8 GHz (marked by the black dashed line), $\varepsilon = 6.5 + 0.18i$, $\mu_x = 1$ and $\mu_y = -2.3$. Under these parameters, we plot the IFC of this HMM, as is shown in Fig. 9(b). The IFC is a general hyperbola in which the two asymptotes are represented by red dashed lines. Because the density of states along the directions of two asymptotes is largest, inside the general HMM the energy is mainly confined on two pathways along the directions of asymptotes [10]. Once we put a dark atom on the pathway, the coupling of the bright and dark atoms can be maintained at a long distance. Now we change the ENP-HMM in Fig. 4 into the general HMM given in Fig. 9(b) and change the value of $C_2$ into 1.25 pF. Moreover, two same wide SRRs are placed on two pathways just for enhancing the coupling. From the simulated reflectance spectrum in Fig. 9(c), long range EIT indicated by the purple arrow is also realized. In Fig. 9(d) we give the 2D electric field distribution at 0.8 GHz. It is seen that the energy is transferred from the bright atom to the dark atoms due to EIT effect. In the general HMM, the energy mainly propagates along the directions of two asymptotes, which are indicated by the white dashed lines. Of course, comparing Fig. 7 with Fig. 9, we find that the ENP-HMM used in the long range EIT is more effective than the general HMM.

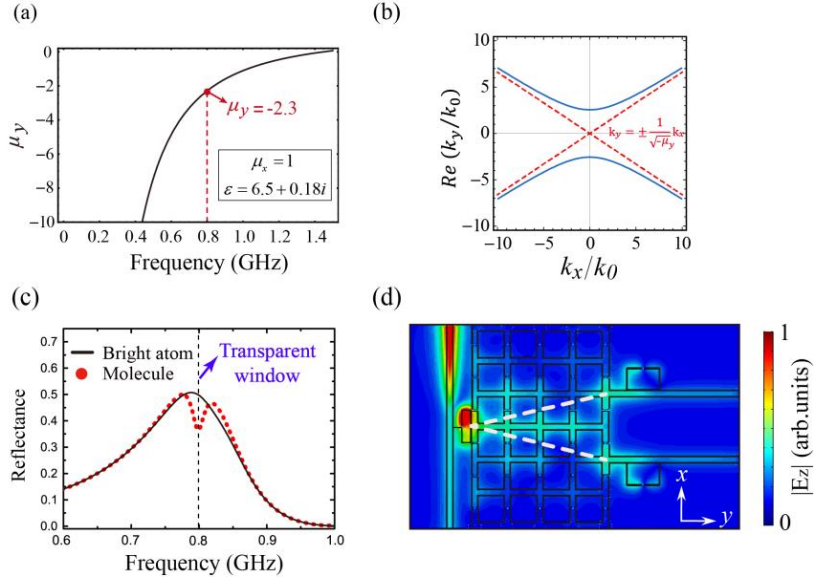

Fig. 9. (a) Effective parameters based on the TLs. The reference frequency is marked by the black dashed line. (b) The IFC of the designed general HMM is a standard hyperbola, where the asymptote represented by red dashed lines. (c) Simulated reflectance spectra of a single bright atom (black solid line) and a molecule (red dotted line) in the general HMM, respectively. The position of transparent window is marked by the purple arrow. (d) Simulated 2D electric field of the molecule at 0.8GHz. The main pathways of energy in the general HMM are marked by white dashed lines.

*B. Ordinary and extraordinary modes in the HMM*

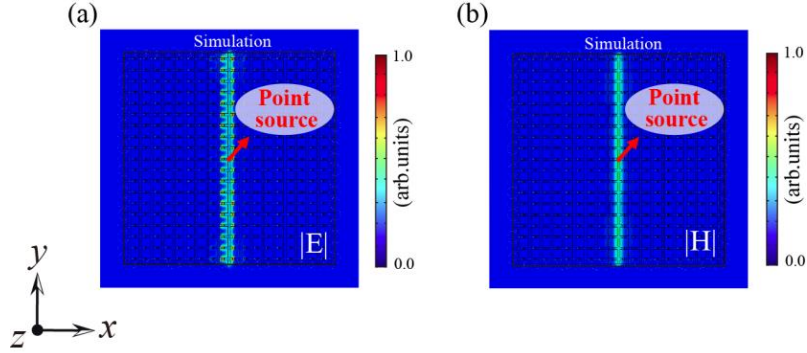

Fig. 10. Collimated electric field (a) and magnetic field (b) pattern when a point source is put at the center of the HMM. The parameters of the HMM are the same as those in Fig. 3(b) for $\mu_y = -81$.

Because of the anisotropy of permittivity or permeability, both ordinary and extraordinary modes would be excited in the HMM. For a long coupling distance, the near-field coupling contribution from ordinary modes is negligible. However, for a short coupling distance, both ordinary and extraordinary modes will play roles in the establishment of EIT. In our physical model, we have considered the near-field coupling contributed by the ordinary modes and the converted-far-field coupling contributed by the extraordinary modes in the HMM. Nevertheless, the contribution of ordinary modes can be ignored for a quasi-static TE-polarized solution of 2D microwave transmission lines loaded with lumped elements. We now explain this point. Considering TL-based HMMs in which the parameters are the same as those in Fig. 3(b). For the TE-polarized extraordinary modes, when $\varepsilon = 6.5 + 0.18i$, $\mu_x = 1$ and $\mu_y = -81$, the dispersions are very flat lines, as are shown by the red lines in Fig. 3(c).

At the same frequency, for the TM-polarized ordinary modes, the dispersion is a circle. However, similar to [7], in this kind of TLs, the excited waves are mainly TE-polarized modes. To check this point, in Fig. 10, we put a point source at the center of the TLs and simulate the emission patterns. In Fig. 10, we only see a collimated pattern that corresponds to the dispersion of extraordinary modes. We do not see the pattern propagating into all directions that corresponds to the dispersion of ordinary modes. Therefore, the TE-polarized extraordinary modes dominate in our designed HMM. The impact of TM-polarized ordinary modes in our system can be neglected.

**Funding**

This work is supported by the National Key Research Program of China (No. 2016YFA0301101), the National Natural Science Foundation of China (NSFC) (Grant Nos. 11774261, 11474220, 11234010 and 61621001).

**Acknowledgment**

GSA thanks the Bio Photonics initiative of the Texas A & M University for support.